# Ten Quick Tips for Harnessing the Power of ChatGPT/GPT-4 in Computational Biology

Tiago Lubiana[1,*], Rafael Lopes[2], Pedro Medeiros[3], Juan Carlo Silva[1], Andre Nicolau Aquime Goncalves[4], Vinicius Maracaja-Coutinho[5,6,7,8,9], Helder I Nakaya[1,10*]

[1] School of Pharmaceutical Sciences, University of São Paulo, São Paulo, Brazil
[2] Department of Epidemiology of Microbial Diseases and Public Health Modeling Unit, Yale School of Public Health, New Haven, CT, USA.
[3] TauGC Bioinformatics, São Paulo, Brasil
[4] Oxford Vaccine Group, University of Oxford, Oxford, United Kingdom
[5] Advanced Center for Chronic Diseases, Universidad de Chile, Santiago, Chile
[6] Centro de Modelamiento Molecular, Biofísica y Bioinformática - CM2B2, Facultad de Ciencias Químicas y Farmacéuticas, Universidad de Chile, Santiago, Chile.
[7] ANID Anillo ACT210004 SYSTEMIX, Rancagua, Chile.
[8] Anillo Inflammation in HIV/AIDS - InflammAIDS, Santiago, Chile.
[9] Beagle Bioinformatics, São Paulo, Brasil & Santiago, Chile
[10] Hospital Israelita Albert Einstein, São Paulo, Brazil
* Corresponding authors

Email:

tiago.lubiana.alves@usp.br

hnakaya@gmail.com

## Introduction

The rise of advanced chatbots, such as ChatGPT, has stirred excitement and curiosity in the scientific community. Powered by large language models (LLMs) GPT-3.5 and GPT-4, ChatGPT is a General Purpose Technology with the potential to impact the job market and research endeavors in numerous fields [1]. Although similar models have been fine-tuned for biology-specific projects, including text-based analysis and biological sequence decoding [2,3], ChatGPT provides a natural interface for bioinformaticians to begin using LLMs in their activities. This tool is already accelerating various activities undertaken by computational biologists, ranging from data cleaning to interpretating results and publishing.

However, with great power comes great responsibility. As scientists, we must harness the full potential of ChatGPT while adhering to ethical guidelines and avoiding pitfalls associated with the technology.

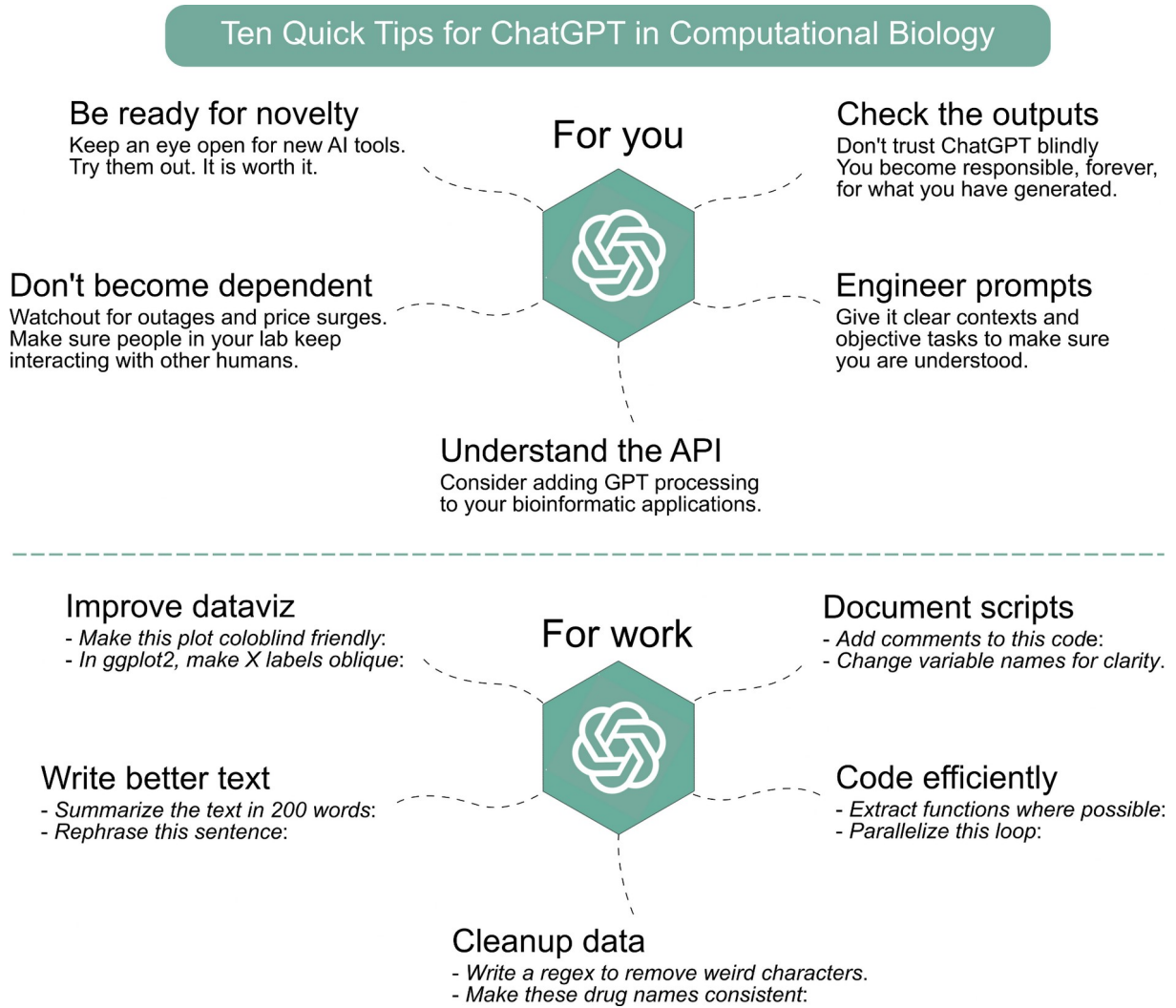

**Figure 1: Ten Quick Tips for ChatGPT in Computational Biology.** The tips are categorized into five mindset and study suggestions and five practical tips, each with simplified but effective prompt suggestions.

Here, we provide ten insightful tips designed to help computational biologists optimize their workflows with ChatGPT, ranging from basic prompts to more advanced techniques. Although our primary focus is on the current ChatGPT/GPT-4 model, we believe that these tips will remain relevant for future iterations of the technology, as well as other LLMs and chatbots (such as Meta's LLaMa and Google's Bard)[4][5]. We invite you to explore our ten tips (summarized in Fig 1) aimed at effectively utilizing ChatGPT to advance computational biology research while maintaining a strong commitment to research integrity.

**Tip 1. Embrace the Technology and Be Ready for Novelty**

ChatGPT, a powerful tool for coding and academic writing tasks, is rapidly gaining traction in the scientific community. While exercising critical judgment and not blindly accepting everything it produces is important, incorporating ChatGPT into your workflow can undoubtedly improve efficiency. We echo van Dis and colleagues' recommendation that every research group should immediately explore and discuss the potential uses of chatbots for their work[6].

Chatbot technology is evolving very fast. Although our tips will be valuable in the near future, new tools and applications are emerging every day. As we finalize this manuscript, ChatGPT has introduced support for plugins and a new partnership with Wolfram Alpha, significantly extending its mathematical and computational capabilities[7]. Thus, one of the most valuable tips we can offer is to be prepared for novelty and remain open to testing new AI advances.

The speed and quality improvements introduced by these novelties are rapidly changing the way we work. [1] By embracing technology, you can increase your changes in the job market and in competitive academic settings. In other words, while ChatGPT will not replace computational biologists, it is likely that researchers who do not use it (and similar tools) will lag behind in competitiveness.

**Tip 2. Improve Code Readability and Documentation**

Programming is a central skill of computational biologists. However, code outputs in academia, such as software, packages, web applications, and analysis scripts are written by time-constrained students and postdocs. Often, these codes do not follow industry-level best practices [8] and require some cleaning and better documentation [9]. Nevertheless, these pieces of code work fine in practice - we just generally wish they were more readable.

Thus, a good starting point to begin harnessing the power of ChatGPT is to make your favorite scripts more readable. Simple prompts such as "*Add explanatory comments to this code:*" or "*Rename the variables for clarity:*" can already do wonders for future readers of the code. ChatGPT can also help document functions by generating full *roxygen2* syntax in R and docstrings in python, inferring meaning from variable names and code logic. A sample prompt to start documenting can be "*Render roxygen2 documentation for the function:*".

**Tip 3. Write Code Efficiently**

In addition to improving the appearance of the code, ChatGPT can be of great help in constructing the logic of scripts. Bioinformatics settings are diverse, and computational biologists often act as jacks of all trades, handling multiple analyses across collaborations. ChatGPT accelerates the learning of new tools, as it provides an interactive environment capable of commenting on different parts of pipelines. It can provide reasonable code chunks on demand and help fix errors by simply copying and pasting error messages into the dialog [10][11]. Of course, expert humans should review the newly produced code and prevent any semantic error (see Tip. 7).

Furthermore, ChatGPT can perform several functional refactorings. Prompts such as "*Extract functions for increased clarity:*" or "*Re-write and optimize this for loop:*" can improve code modularity and even save computational resources. When refactoring, it is important to set up good tests to prevent introducing bugs [12]. While ChatGPT can also help you with setting up testing infrastructure (with prompts like "*Write a unit test for the following function and help me implement it:*"), it is crucial to double-check what it generates to ensure it is covering what it should.

A middle ground between using ChatGPT and implementing full-scale LLM applications is to add ChatGPT to integrated development environments (IDEs) via plugins. For example, it is currently possible to use GPT-3.5 and GPT-4 in Visual Studio Code (VSCode) and open-source plugins are available (https://github.com/gencay/vscode-chatgpt). For the bioinformaticians using R and RStudio, there are options such as *gptstudio* (https://github.com/MichelNivard/gptstudio).

**Tip 4. Use ChatGPT to Enhance Data Cleanup**

In addition to writing scripts, computational biology research involves cleaning and reconciling data, ensuring it is consistent and free of errors before running the analysis. Data and metadata come in various formats, and while ChatGPT will not identify outliers or fix missing data, it can suggest tools for most common tasks and provide code snippets. It can also partner up with Excel, offering guidance and writing macros [13].

As expected, ChatGPT proves most useful when processing datasets with natural language entries. If you manage a database or re-analyze public datasets, you likely have to deal with inconsistent input entered by submitters. While the current tool cannot consistently match data to unique identifiers (such as those provided by databases or ontologies [14]), it can add more consistency and facilitate manual or automatic biocuration steps [15]. A clear application is to write regular expressions given a few examples, with prompts such as "*Write me regex for R/python/Excel with a pattern that will extract {} from {}*".

ChatGPT can greatly help in normalizing labels directly and executing human-like complex natural language cleanups, like those found in open-field formularies. For small datasets, you can clean up data directly in the ChatGPT interface, with prompts such as "*Act as a table. Add a new column with consistent labels to this dataset:*". For larger applications, one can use add-ons, such as GPT for Google Sheets (https://gptforwork.com/), or even write code that uses the API directly (see Tip 9).

**Tip 5. Use ChatGPT to Improve Your Data Visualization**

Data visualization is an essential component of computational biology research, and ChatGPT can be a valuable tool to assist in creating effective and informative figures. One remarkable capacity of this tool is its proficiency in popular visualization libraries, such as *ggplot2* and *matplotlib* (e.g. "*Create a ggplot2 violin plot with a log10 Y axis*"). This expertise enables it to assist users in overcoming syntax challenges, suggesting new visualization techniques, and enhancing existing figures.

Image-parsing by GPT-4 has been announced, but as of the time of writing is not yet available for common users. [16]Thus, while we may soon be able to get direct feedback on images, we can still leverage GPT-4's ability to parse code for plotting and receive valuable guidance on areas for improvement. For example, ChatGPT can help you choose appropriate colors for your figures, make the figures more accessible for color-blind individuals, and suggest ways to improve the layout of your visualizations. A practical example of a prompt that can lead to meaningful improvements in your visualizations is asking ChatGPT to *"Change my code to make the plot color-blind friendly"*.

It's important to note that ChatGPT's suggestions should be used as a starting point for further exploration and refinement, as good figure design involves careful consideration of data, layout, and style. To make the most of ChatGPT's capabilities, it is essential to familiarize oneself with the principles of good figure design, which can be found in resources such as the PLOS Computational Biology article "Ten Simple Rules for Better Figures" [17]. Overall, by harnessing ChatGPT's potential in generating and refining visualizations, computational biologists can enhance their research output, create more accessible figures, and communicate their findings more effectively.

**Tip 6. Use ChatGPT to Improve Your Writing**

While AI-assisted writing in science has been steadily growing [18], ChatGPT has made this technology accessible to a much wider range of scientists and researchers. One of the most valuable features for authors, especially non-native English speakers, is its aid in expressing ideas more clearly. Clear and effective communication is especially important in computational biology, where experts must be capable of conveying complex ideas to colleagues with varying scientific backgrounds, using language that is understandable by mathematicians, biologists, and computer scientists alike. ChatGPT improves the clarity of text, by providing new ways of ordering thoughts, with prompts like "Provide me some different versions of the following sentence:".

ChatGPT can also help with reformatting text and summarizing thoughts, with prompts such as "Summarize this text in a 200-word conference abstract:". Although it will rarely produce an output that you will fully like, it can break the initial barrier, helping to overcome writer's blocks. It can do so also by helping outline documents, from papers to teaching plans, both by creating bulleted lists from natural language and by converting bulleted lists into a final format.

Besides scientific writing, ChatGPT can be utilized for several other writing tasks, such as creating emails, grant reports, tutorials, and documentation (see Tip 2), and selecting appropriate keywords for publications. Furthermore, it can modify the text to cater to various readerships, including composing media releases, simplifying research for non-specialists, or adapting language from a biologist-based audience to a computer-science-based one.

Regardless of where you use ChatGPT to improve your writing, be sure to disclose its usage (or other language models) as a writing tool to prevent any misunderstandings.[19] Guidelines for responsible usage are emerging regarding the ethical use of chatbots as writing aids, particularly in the context of publishing manuscripts. [20,21] We advise researchers to familiarize themselves with the discussions and check publisher guidelines whenever using ChatGPT for publishable research.

**Tip 7. Ensure You Understand - or Know How to Test - What it Generates**

While ChatGPT can be a powerful tool for writing code and text in computational biology pipelines, it's important to be careful when applying it to complex analysis. In some cases, ChatGPT may hallucinate or add bugs that can produce silent errors and lead to false conclusions.

For beginners in computational programming, the suggestion of functions or libraries that do not exist can be a significant hurdle and reinforces the need for human intervention. Therefore, it's important to study tutorials provided by developers and publications related to the topic of interest. When using ChatGPT to help with syntax, it's crucial to only ask for help with syntax that you have already studied and can understand - or at least test - the results.

A similar caution should be applied when using ChatGPT for writing articles or interpreting results. Double-check what you read, understand, and agree with everything the chatbot has generated. In the end, you will be responsible for the text, not OpenAI or ChatGPT.

**Tip 8. Learn the Basics of Prompt Engineering/Design**

Being an emerging field, the terms are still being discussed, but the importance of knowing how to interact with a non-deterministic system aiming for an objective result is vital. Prompt engineering/design involves crafting prompts that effectively communicate, examples, personas, and goals, to generate response templates that fit your objectives [22,23]. It is also important to set evaluation metrics to feed the model toward more assertive results within the limits of available tokens.

A good example of a prompt is: *"ChatGPT, I'd like to learn about the use of GATK tools in bioinformatics. Could you provide a brief overview of GATK, its main applications, and some popular tools within the GATK suite that are commonly used in the field of bioinformatics? Please include any advantages and limitations associated with these tools."* This prompt is effective because it clearly states the context (bioinformatics), specifies the topic (GATK tools), outlines the desired information (overview, applications, popular tools, advantages, and limitations), and provides a concise and focused question for the AI to address.

In contrast, a bad example would be *"Tell me about GATK."* This prompt is ineffective because it lacks context (no mention of bioinformatics), is vague about the topic (just mentioning GATK, not specifically GATK tools), doesn't specify desired information (no details about what aspects of GATK to discuss), and provides an overly broad and open-ended question, which may result in less relevant or less focused responses.

By providing more context, details, and specific goals, the good example is more likely to generate a relevant and informative response from ChatGPT, while the bad example may lead to a less satisfying outcome. The addition of new parameters after the first outputs for the refinement is an open possibility, yet caution must be exercised as the risk of loss of context increases as dialogues become longer, subtle, and more complex. As such, it is imperative to prioritize specificity, objectivity, and completeness in initial interactions to mitigate the potential for hallucinations and deviations.

**Tip 9. Consider the GPT API to Extend Your Applications**

In addition to using the graphical interface, OpenAI's API allows fine-tuning GPT to better fit your work. You can use the API to improve interfaces for user-friendly applications, allowing the user to interact with your software using human language and have GPT convert it into executable code. The API can also be part of pipelines on your own workflow. For instance, in a text mining and tokenization pipeline, it can be used to extract entities from the text database or to summarize text based on desired stopwords.

Fine-tuning involves the manipulation of four parameters that modulate the creativity of the system: *temperature*, *top_p*, *frequency_penalty*, and *presence_penalty*. The *temperature* and *top_p* parameters control the degree of boldness and non-determinism exhibited in the output, and high values reduce the repetitiveness of responses in terms of content and meaning. The *frequency_penalty* and *presence_penalty* parameters regulate the likelihood of token (word) repetition in the output, and higher values of these parameters minimize repeated tokens. Note that reproducibility is not guaranteed even when fixing parameters, as GPTs are non-deterministic. Nevertheless, fine-tuning can potentially result in cleaner, less repetitive, and more concise outputs.

The API can also help when input contains text larger than allowed in web prompts (around 4,000 characters). Large documents can be parsed with GPT by employing tools such as LangChain (https://github.com/hwchase17/langchain), which are capable of modifying extensive documents from diverse sources for access by the model and facilitating responses in a more organized manner.

However, this field is evolving rapidly, and developers are working swiftly to incorporate the model with tools that address its limitations. New features must be promptly available to keep up with the accelerated pace of advancement.

**Tip 10. Don't Become Too Dependent on ChatGPT**

While ChatGPT is a game changer, it is important to remember that it is still in the early stages of development. While it may seem like a magic bullet for many researchers, there are still some issues that need to be considered. It's essential not to become too dependent on ChatGPT and to have backup plans in place, remembering how to do things "by hand" when necessary.

One of the key challenges of ChatGPT is that it is being tested to the limit, and the platform has experienced shutdowns and outages recently. This can be especially problematic for researchers who rely heavily on ChatGPT for their work. Moreover, there are currently no commercial alternatives to ChatGPT, and no open-source or non-profit endpoints available. Over-reliance on any single entity may disrupt your scientific workflow and can be particularly difficult for those in the Global South, where price surges can be prohibitive.

If you're a mentor or team leader, it's essential to ensure that your team is not overly dependent on ChatGPT and that they have the support they need to succeed. While ChatGPT is a powerful tool, it should not replace mental health professionals or the social interactions that come from collaborating with coworkers. If ChatGPT is providing help that was previously coming from colleagues, it is important to find alternative ways to foster social interaction, such as coding dojos, pair programming, or social and sports events. Always strive for a balanced approach when using any AI tools, making sure your team continues to develop essential skills and knowledge independently.

**Conclusion**

ChatGPT and other LLM chatbots are powerful tools that are increasingly becoming essential to scientists and programmers, as well as the various other professionals in between. They offer the potential to improve productivity and simplify complex workflows, especially in cases involving repetitive or minor tasks. It pays to invest time in understanding the tool's applicability and limitations and avoid over-reliance.

Keep in mind they are general-purpose tools [23] To keep track of new, creative uses for these tools in bioinformatics, we have set up a GitHub repository to crowd-curate content arising on the matter: https://github.com/csbl-br/awesome-compbio-chatgpt . We believe that these technologies will help computational biologists to perform their activities more efficiently, ultimately improving the pace of scientific discovery. We hope that these tips will help you use ChatGPT to complement (and not substitute) your workflows while remaining aware of the various applications and implications of this technology.

**LLM assistance statement**

GPT-4 and ChatGPT were used for writing, coding, and formatting assistance in this project.

**Funding statement**

T.L. is funded by FAPESP Grant #19/26284-1, J.C.S. is funded by FAPESP Grant #19/27139-5. V.M.C. is funded by FONDECYT-ANID (1211731), FONDAP-ANID (15120011), STIC/AmSud-ANID (STIC2020008) and Anillo-ANID (ACT210004 and ATE220016).

**References**

1. Owens B. How Nature readers are using ChatGPT. Nature. 2023;615: 20.

2. Ferruz N, Schmidt S, Höcker B. ProtGPT2 is a deep unsupervised language model for protein design. Nat Commun. 2022;13: 4348.

3. Thorp HH. ChatGPT is fun, but not an author. Science. 2023;379: 313.

4. Touvron H, Lavril T, Izacard G, Martinet X, Lachaux M-A, Lacroix T, et al. LLaMA: Open and Efficient Foundation Language Models. 2023. Available: http://arxiv.org/abs/2302.13971

5. Bard. [cited 23 Mar 2023]. Available: https://bard.google.com/

6. van Dis EAM, Bollen J, Zuidema W, van Rooij R, Bockting CL. ChatGPT: five priorities for research. Nature. 2023;614. doi:10.1038/d41586-023-00288-7

7. ChatGPT Gets Its "Wolfram Superpowers"!—Stephen Wolfram Writings. [cited 24 Mar 2023]. Available: https://writings.stephenwolfram.com/2023/03/chatgpt-gets-its-wolfram-superpowers/

8. Trisovic A, Lau MK, Pasquier T, Crosas M. A large-scale study on research code quality and execution. Sci Data. 2022;9: 60.

9. Filazzola A, Lortie CJ. A call for clean code to effectively communicate science. Methods Ecol Evol. 2022. doi:10.1111/2041-210x.13961

10. Shue E, Liu L, Li B, Feng Z, Li X, Hu G. Empowering Beginners in Bioinformatics with ChatGPT. bioRxiv. 2023. p. 2023.03.07.531414. doi:10.1101/2023.03.07.531414

11. Sobania D, Briesch M, Hanna C, Petke J. An analysis of the automatic bug fixing performance of ChatGPT. 2023. doi:10.48550/ARXIV.2301.08653

12. Hunter-Zinck H, de Siqueira AF, Vásquez VN, Barnes R, Martinez CC. Ten simple rules on writing clean and reliable open-source scientific software. PLoS Comput Biol. 2021;17: e1009481.

13. Williams KL. Using ChatGPT with Excel. In: Journal of Accountancy [Internet]. 30 Jan 2023 [cited 26 Mar 2023]. Available:

https://www.journalofaccountancy.com/news/2023/jan/using-chatgpt-with-excel.html

14. McMurry JA, Juty N, Blomberg N, Burdett T, Conlin T, Conte N, et al. Identifiers for the 21st century: How to design, provision, and reuse persistent identifiers to maximize utility and impact of life science data. PLoS Biol. 2017;15: e2001414.

15. Amy Tang Y, Pichler K, Füllgrabe A, Lomax J, Malone J, Munoz-Torres MC, et al. Ten quick tips for biocuration. PLoS Comput Biol. 2019;15: e1006906.

16. OpenAI. GPT-4 Technical Report. 2023. doi:10.48550/ARXIV.2303.08774

17. Rougier NP, Droettboom M, Bourne PE. Ten simple rules for better figures. PLoS Comput Biol. 2014;10: e1003833.

18. Hutson M. Could AI help you to write your next paper? Nature. 2022;611: 192–193.

19. Stokel-Walker C. ChatGPT listed as author on research papers: many scientists disapprove. In: Nature Publishing Group UK [Internet]. 18 Jan 2023 [cited 26 Mar 2023]. doi:10.1038/d41586-023-00107-z

20. Stokel-Walker C, Van Noorden R. What ChatGPT and generative AI mean for science. Nature. 2023;614: 214–216.

21. Tools such as ChatGPT threaten transparent science; here are our ground rules for their use. Nature. 2023. p. 612.

22. White J, Fu Q, Hays S, Sandborn M, Olea C, Gilbert H, et al. A prompt pattern catalog to enhance prompt engineering with ChatGPT. 2023. doi:10.48550/ARXIV.2302.11382

23. Beurer-Kellner L, Fischer M, Vechev M. Prompting is programming: A query language for large language models. 2022. doi:10.48550/ARXIV.2212.06094